\documentclass[journal,letterpaper]{IEEEtran}
\usepackage[utf8]{inputenc}
\usepackage[english]{babel}
\ifCLASSINFOpdf
  \usepackage[pdftex]{graphicx}
  \graphicspath{{./figures/}{./plots/}{./bio/}}
  \DeclareGraphicsExtensions{.pdf,.jpeg,.jpg,.png}
\fi
\usepackage[caption=false,font=footnotesize]{subfig}
\usepackage[export]{adjustbox}
\usepackage[cmex10]{amsmath,mathtools}
\usepackage[linesnumbered,lined,ruled]{algorithm2e}
\usepackage[usenames,dvipsnames,svgnames]{xcolor}
\usepackage{array,booktabs,url,units}

\begin{document}

\title{Smarter Cities with Parked Cars as Roadside Units}
\author{Andre~B.~Reis, Susana~Sargento, and~Ozan~K.~Tonguz
\thanks{
A.~B.~Reis is with the Department of Electrical and Computer Engineering, Carnegie Mellon University, Pittsburgh, PA 15213-3890 USA, and also with the Universidade de Aveiro, Instituto de Telecomunicacoes, Aveiro 3810-193, Portugal (e-mail: abreis@cmu.edu).
}
\thanks{S.~Sargento is with the Universidade de Aveiro, Instituto de Telecomunicacoes, Aveiro 3810-193, Portugal (e-mail: susana@ua.pt).
}
\thanks{O.~K.~Tonguz is with the Department of Electrical and Computer Engineering, Carnegie Mellon University, Pittsburgh, PA 15213-3890 USA (e-mail: tonguz@ece.cmu.edu).
}
}

\maketitle

\begin{abstract}
Real-time monitoring of traffic density, road congestion, public transportation, and parking availability are key to realizing the vision of a smarter city and, with the advent of vehicular networking technologies such as IEEE~802.11p and WAVE, this information can now be gathered directly from the vehicles in an urban area. To act as a backbone to the network of moving vehicles, collecting, aggregating, and disseminating their information, the use of parked cars has been proposed as an alternative to costly deployments of fixed Roadside Units.

In this paper, we introduce novel mechanisms for parking vehicles to self-organize and form efficient vehicular support networks that provide widespread coverage to a city. These mechanisms are innovative in their ability to keep the network of parked cars under continuous optimization, in their multi-criteria decision process that can be focused on key network performance metrics, and in their ability to manage the battery usage of each car, rotating roadside unit roles between vehicles as required. We also present the first comprehensive study of the performance of such an approach, via realistic modeling of mobility, parking, and communication, thorough simulations, and an experimental verification of concepts that are key to self-organization. Our analysis brings strong evidence that parked cars can serve as an alternative to fixed roadside units, and organize to form networks that can support smarter transportation and mobility.

\end{abstract}

\newcommand*{\Comb}[2]{{}^{#1}C_{#2}}
\newcommand{\sol}{\mathcal{S}}

\newcommand{\asig}{ \mathrm{a_{sig}} }
\newcommand{\asat}{ \mathrm{a_{sat}} }
\newcommand{\acov}{ \mathrm{a_{cov}} }
\newcommand{\abat}{ \mathrm{a_{bat}} }

\newcommand{\wsig}{ {w_\mathrm{sig}} }
\newcommand{\wsat}{ {w_\mathrm{sat}} }
\newcommand{\wcov}{ {w_\mathrm{cov}} }
\newcommand{\wbat}{ {w_\mathrm{bat}} }

\newcommand{\SCM}{\mathrm{\textsc{scm}}}
\newcommand{\LM}{\mathrm{\textsc{lm}}}
\newcommand{\LMC}{\mathrm{\textsc{lmc}}}
\newcommand{\LMS}{\mathrm{\textsc{lms}}}
\newcommand{\BAT}{\mathrm{\textsc{bat}}}
\newcommand{\scm}{\mathrm{scm}}
\newcommand{\lm}{\mathrm{lm}}
\newcommand{\lmc}{\mathrm{lmc}}
\newcommand{\lms}{\mathrm{lms}}

\newcommand{\sqkm}{km\textsuperscript{2}}
\newcommand{\sqm}{m\textsuperscript{2}}

\definecolor{signal2}{HTML}{FD7400}
\definecolor{signal3}{HTML}{FFE11A}
\definecolor{signal4}{HTML}{BEDB39}
\definecolor{signal5}{HTML}{1F8A70}
\definecolor{histbar}{HTML}{0074D9}

\newcommand\crule[3][black]{\textcolor{#1}{\rule{#2}{#3}}}
\newcommand{\signalDistKey}{
\raisebox{-.05\height}{\crule[signal2]{2mm}{2mm}}\,{\footnotesize 2}\quad
\raisebox{-.05\height}{\crule[signal3]{2mm}{2mm}}\,{\footnotesize 3}\quad
\raisebox{-.05\height}{\crule[signal4]{2mm}{2mm}}\,{\footnotesize 4}\quad
\raisebox{-.05\height}{\crule[signal5]{2mm}{2mm}}\,{\footnotesize 5}
}

\newcommand{\histKey}{
\raisebox{-.10\height}{\crule[histbar]{2mm}{2mm}}\,{\footnotesize density}
}

\newcommand{\tabshortstack}[1]{{\bgroup\renewcommand{\arraystretch}{0.8}\begin{tabular}[t]{@{}r@{}}#1\end{tabular}\egroup}}
\newcommand{\tabshortcstack}[1]{{\bgroup\renewcommand{\arraystretch}{0.8}\begin{tabular}[c]{@{}r@{}}#1\end{tabular}\egroup}}
\newcommand{\tabmedstack}[1]{{\bgroup\renewcommand{\arraystretch}{1.2}\begin{tabular}[c]{@{}c@{}}#1\end{tabular}\egroup}}


\section{Introduction} 
\label{sec:introduction}
\IEEEPARstart{A}s the world population continues to shift from rural to urban areas, smarter transportation becomes an increasingly necessary part of urban development. With urbanization comes a greater need for mobility, both in the form of personal vehicles and of public transportation, and consequently a demand for better transit and parking information. A greater awareness of available transportation resources is therefore a key element of a Smart City.

The advent of vehicular networking, through the IEEE 802.11p and WAVE standards, allows individual vehicles to communicate and report their activity, which enables the direct monitoring of transportation resources. Traffic density and road congestion can be measured in real-time~\cite{alam2011,lan2008,felice2014}, parking space availability estimated from cars that arrive and depart~\cite{caliskan2006,panayappan2007}, and the location of buses and other public transport can be tracked directly and efficiently~\cite{wang2002}. The envisioned traffic efficiency applications assume one such infrastructure deployment in the form of Roadside Units (RSUs), but the prohibitive costs associated with RSU hardware, installation, and maintenance have severely limited their adoption~\cite{michiganstudy,dotcosts}.

Vehicles that park in urban areas can be leveraged to take on the roles of fixed roadside units, positioning themselves as an effective alternative to costly deployments of network infrastructure. Research shows that parked cars equipped with 802.11p technology are able to self-organize and create a vehicular support network in an urban area, replacing or augmenting existing roadside units~\cite{vtc2015,tits2017}. Such a vehicular support network, created from parked cars, is able to serve an important role in achieving the vision of a Smarter City: the ubiquity of parked cars in the urban area allows them to form widespread mesh networks, while their On-Board Unit (OBU) hardware lets them be aware of nearby vehicle activity and monitor the city's transportation resources.

This paper introduces a new approach for parked car self-organization. This approach advances existing techniques in several ways: the proposed mechanisms optimize networks of parked cars beyond their initial grouping, which leads to a more efficient selection of cars to act as RSUs; a multi-criteria decision making process acts directly on key metrics of signal strength, RSU saturation, and network coverage, allowing for a precise control of the resulting support network; and car battery usage is factored into the decision process, with RSU roles being rotated among parked cars, ensuring a controlled use of each vehicle's battery resources.

The work presented in this paper also includes the first comprehensive evaluation of the concept of leveraging parked cars as effective roadside unit replacements. Thorough simulation studies analyze how parked cars initially organize to form new networks, and how those networks behave in their steady-state. We shed light on the number of parked cars that need to be recruited to provide urban coverage in various scenarios, and analyze the quality and strength of the resulting networks in detail.
Upper and lower bounds are established for the balance between better coverage strength and fewer roadside units, providing a unique frame of reference to judge the performance of self-organizing approaches, 
and various iterations on the design of the newly introduced decision processes are all seen to operate near the most optimal bounds. 
We bring in detailed, realistic models of the distribution of parking events and their respective duration~\cite{smartvehicles16}, and validate the proposed mechanisms against them.
Finally, this work presents the first empirical study of the process through which vehicles learn their own coverage map by overhearing beacons from other moving cars, an essential element to the self-organization process, and prove its validity. 
With this, we provide strong confirmation that self-organizing approaches are suitable for the creation of vehicular support networks from parked cars, and provide both an efficient set of new decision algorithms to do so and a comprehensive simulation platform to evaluate their performance on.

\clearpage
This work's main contributions can be summarized as follows:
\begin{itemize}
	\item A new approach for better parked car self-organization is introduced. Instead of relying on individual decisions, we give newly parked cars the role of decision makers, allowing them to reconfigure RSUs in their vicinity. This ensures a continued optimization of the resulting network. 
	\item A multi-criteria decision making solution is formulated for the problem of assigning RSU roles to available cars.
	\item A system to limit the battery power draw on each car is developed, allowing RSU roles to be rotated between available vehicles as a part of each decision step. 
	\item An extensive study is provided for the performance of a self-organized approach for parked cars. We show, for the first time:
	\begin{itemize}
		\item how networks of parked cars come together;
		\item the quality of the support network that is created;
		\item what the best balance between the number of recruited cars and the quality of the network is;
		\item the sensitivity to the frequency of parking events, and to the radio range of the vehicles' hardware;
		\item the performance of the network under realistic parking behaviors drawn from empirical data.
	\end{itemize}
	\item An experimental study is shown for the process that cars employ to learn their own map of coverage, and its resilience is demonstrated.
\end{itemize}



The remainder of this paper is organized as follows. Section~\ref{sec:a_new_approach} gives an overview of the main concepts behind vehicular support networks that consist of parked cars, and describes our new approach for self-organization based on local decision makers. Section~\ref{sec:evaluation} presents a thorough evaluation of the proposed decision mechanisms, their efficiency, and their performance in realistic scenarios. An empirical study on the learning process that newly parked cars use to determine their coverage is shown in Section~\ref{sec:empirical}. Related work is presented in Section~\ref{sec:relatedwork}, and finally, concluding remarks are given in Section~\ref{sec:conclusion}.



\section{Improving Decisions for Self-Organized Networks of Parked Cars} 
\label{sec:a_new_approach}
\begin{figure}[t]
	\includegraphics[trim=80mm 90mm 80mm 90mm,clip,frame,width=\linewidth]{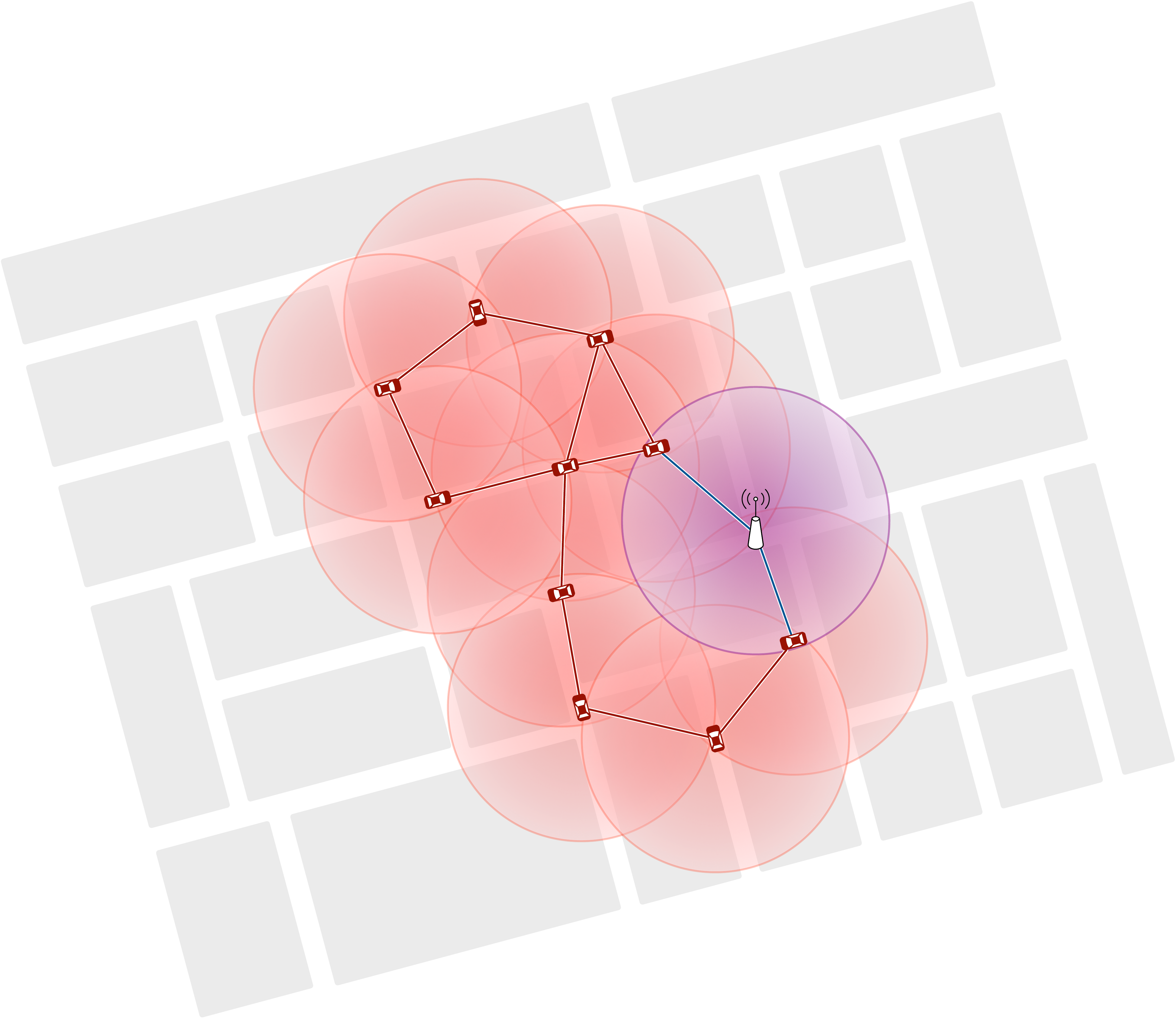}
	\caption{Parked cars can act as relays for existing roadside units, and also form mesh networks to extend the area covered by the vehicular support network.}
	\label{fig:modesofoperation}
\end{figure}

Groups of parked cars that self-organize can become roadside units in a standalone mode, forming a mesh network with point-to-point links between them, or as relays to existing fixed RSUs, extending their coverage range. These modes can also coexist, as is shown in Figure~\ref{fig:modesofoperation}. The masses of parked cars must then work towards self-organization while meeting a number of goals: providing widespread coverage to the urban area; selecting, from large pools of parked vehicles, which should take on RSU roles; and managing the battery power drain on those cars where the On-Board Unit is kept active.

The work in~\cite{tits2017} introduced a number of novel concepts that facilitate this self-organization between cars. Specifically, it proposed a cellular division of the urban area, aligning cell boundaries to GPS coordinates (with a 1~GPS second interval). With this cell division as a common reference, parked cars can then be instructed to listen to Cooperative Awareness Messages (CAMs) from neighboring vehicles, and to build their own coverage maps from these beaconing messages, marking cells where beacons were received from with the signal strength of those messages. 
The exchange of coverage maps between cars allows decision processes to know the utility of each car to the network, and optimize using that information. 

The mechanisms presented in the following section make use of these core concepts. We advance existing work by introducing a completely new approach for the decision process, one where newly parked cars are given the authority to make decisions for their neighbors, assigning and revoking RSU roles to other neighbor RSUs to optimize the local network. Additionally, we formulate the decision problem as a Multiple-Criteria Decision-Making (MCDM) problem that explicitly evaluates key metrics of the resulting network of parked cars, and factors battery usage data from active RSUs into each decision.

\subsection{Local Decision Maker} 
\label{sub:local_decision_maker}
As a new vehicle parks, it first constructs its own coverage map by listening to beacons being sent from moving cars, and gathers the coverage maps of its neighboring RSUs~\cite{vtc2015}. 
Then, it creates a list of candidate roadside unit entities in its 1-hop neighborhood, consisting of itself and other active parked cars. From this pool, the decision maker lists the possible combinations obtained from assigning or revoking RSU roles to the entities in the pool, and assigns a score to each combination. We name a combination of active/inactive entities a \emph{Coverage Solution}, and designate it by $\sol_k$ (where $k$ is the index of the coverage solution).

Once all coverage solutions are ranked, the one with the highest score is applied to the network. To do this, the newly parked car, acting as a authoritative decision maker, sends specially crafted messages reassigning RSU roles in its vicinity. Through this approach, the network of RSUs sees a local optimization every time a new vehicle parks.


\subsection{Ranking Coverage Solutions} 
\label{sub:ranking_coverage_solutions}
A self-organizing approach for parked cars has a number of core goals: the network of cars should aim to provide widespread coverage to the urban area in all of its locations, whenever possible, and to supply that coverage with strong signal, so that moving nodes can connect to the RSU network reliably and with high data rates. It must also aim to minimize the number of parked cars that take on RSU roles, and the time those vehicles spend active as RSUs, drawing power from their respective batteries. 

Because the network is intended to be self-organized, there is an implicit assumption that a central controller is not available, and so decisions must be taken by the parked vehicles themselves, communicating only between them as needed. Poor decisions may lead to either (a), more parked cars being assigned RSU roles than necessary, creating a more crowded network and wasting battery energy, or (b), an RSU network that is suboptimal in signal strength, area of coverage, or both.

To score a coverage solution, we resort to dimensionless analysis: a Weighted Product Model (WPM), where the utility of each solution is determined by multiplying a series of attributes, each of which is raised to the power of a coefficient that represents its weight to the solution's score. This approach eliminates any units of measure, which is ideal to our multi-dimensional decision-making problem. The score of a coverage solution $\mathcal{S}_k$ is, then, a product of attributes $a_{j,k}$, each weighted by a coefficient $w_j$:
\begin{equation}
	\mathrm{score}\left(\sol_k\right) = \prod_{j=1}^{n} \left( a_{j,k} \right)^{w_j}
\end{equation}

Our scoring function integrates the following four attributes:
\begin{itemize}
	\item \textbf{Signal Strength} ($\asig,\wsig$): An average of the best signal strength available at each cell. It is the main attribute that pushes towards a city-wide network with strong coverage.
	\item \textbf{RSU Saturation} ($\asat,\wsat$): The counterweight to signal strength, roadside unit saturation represents the number of RSUs that provide coverage to a cell. Higher average signal strength in the city demands higher RSU saturation and, therefore, a larger number of RSUs.
	\item \textbf{Area of Coverage} ($\acov,\wcov$): The span of the coverage being provided by the RSU network. A strong signal strength does not imply widespread coverage, as it only considers cells where coverage is provided. This attribute can favor solutions with wider coverage.
	\item \textbf{Energy Usage} ($\abat,\wbat$): A measure of how much energy active RSUs in a coverage solution have expended, and how close they are to a maximum threshold. This attribute pushes towards coverage solutions that remove aging RSUs from the network, keeping the battery utilization of active parked cars under control.
\end{itemize}

Each of these attributes is calculated on the expected outcome of a coverage solution, i.e., on the RSU network that would theoretically result if the coverage solution were applied, assigning and revoking RSU roles.


\subsection{Constraining the Search Space} 
\label{sub:constraining_the_search_space}
For a decision maker to score all coverage solutions that may be possible in its neighborhood, it needs to search a space of $2^N$ possible combinations, with $N$ being the number of neighbors plus the decision maker itself. For denser networks of RSUs, the number of coverage solutions in the search space can become too unwieldy and computationally expensive. 

We adopt a more restrained approach, by restricting the search space to coverage solutions that do not revoke more than two RSU roles in the decision maker's neighborhood. Because the self-organized RSU network is grown incrementally as each new car parks, RSUs seen by a decision maker are the result of earlier, equally-optimized decisions, so coverage solutions that revoke higher numbers of roles are unlikely to lead to high decision scores, and it is therefore safe to exclude them from the search space.

Under these constrains, valid coverage solutions will belong to one of three possible categories:
\begin{enumerate}
	\item \textbf{No entities are disabled:} the candidate parked car becomes an RSU, and all neighboring RSUs remain active. A single such solution exists $\left(\Comb{n}{n}=1\right)$, which we denote as $\sol_0$. It increases the number of RSUs in the city by one.
	\item \textbf{One entity is disabled:} either the newly parked car or a neighboring RSU. $\Comb{n}{n-1}$ such solutions exist, and they do not change the number of RSUs in the city. The solution that only disables the newly parked car is a `\emph{no-action}' solution, leaving the network unchanged. We denote this specific case as $\sol_{\textsc{na}}$.
	\item \textbf{Two entities are disabled:} the parked car takes on an RSU role and replaces two existing RSUs; or, both the parked car and a neighbor RSU are disabled. $\Comb{n}{n-2}$ such solutions exist, and they decrease the number of RSUs in the city by one.
\end{enumerate}
With this approach, a decision maker with, e.g., 8 neighbors has to evaluate only 37 solutions of the 256 that were possible.


\subsection{Attribute Models} 
\label{sub:attribute_models}
We now show how to quantify each of the four attributes that are part of the scoring function. Our aim here has been to keep these models as straightforward as possible, so that our subsequent analysis can show, with clarity, what each attribute contributes towards the resulting network of parked cars. An algorithm to calculate all attributes and produce a final decision score is shown afterwards, in Section~\ref{sub:scoring_algorithm}.

\subsubsection{Signal Strength} 
\label{ssub:signal_strength}
Our signal strength attribute is the average signal strength that is expected of a given coverage solution. From the received coverage maps and the vehicle's coverage map, the signal strength of each cell in the resulting solution is taken and averaged.

In this work we use a 1-5 classification that corresponds linearly to a Received Signal Strength Indicator (RSSI). Because the decision algorithm is dimensionless, RSSI may also be used directly. Other metrics such as signal power (e.g., in \unit[]{dBm}) or network bandwidth (e.g., in \unit[]{Mbps}), should be equally practicable, but are not evaluated here. For accuracy, only the cells that the decision maker can cover should be considered, since the vehicle acting as decision maker does not have a complete view of the network beyond these cells.


\subsubsection{RSU Saturation} 
\label{ssub:rsu_saturation}
The RSU saturation attribute reflects the overlap in coverage between active roadside units. For a given cell in a coverage solution, its saturation value equals the number of RSUs that can cover the cell (i.e., the number of RSUs that a vehicle located on the cell will see). This metric reflects the number of RSUs in the area, and also their arrangement: a poor RSU role assignment where RSUs are stacked too close sees more cells with higher saturation (more overlap between the RSUs' coverages), while a more efficient distribution gives a more widespread coverage and more cells with lower saturation.

Saturation is also a measure of redundancy in the RSU network, which may  be desirable from an availability standpoint: when cells are covered by more than a single RSU (saturation greater than $2$), if a car with the RSU role is removed from the network, continued access to the network can be ensured by other RSUs.

As with the signal strength attribute, our saturation attribute is the average of the roadside unit saturation seen at each cell of a coverage solution. Average saturation can range from $[1,\infty[$, but in practice, an efficient algorithm will keep mean saturation in a range of $[1.44, \approx5]$ (see Sections~\ref{sub:balancing_roadside_unit_saturation} and~\ref{sub:decision_performance}).


\subsubsection{Area of Coverage} 
\label{ssub:area_of_coverage}
The signal strength and RSU saturation attributes, on their own, do not ensure a widespread coverage by the network of parked cars, which is also a desired trait for this network. Consider that, once a base network is established, new coverage will be available primarily at the fringes of the covered area, where the signal is weaker, therefore pushing average signal strength down and penalizing coverage solutions that expand the network's reach.
To make sure that local decisions push towards widespread coverage, we include an attribute that quantifies the area of service provided by each coverage solution. 

To compute this attribute, the algorithm begins by counting the number of covered cells in the $\sol_0$ coverage solution, which we defined earlier as the solution that disables no RSUs and, therefore, has the widest possible coverage area (since all other coverage solutions involve disabling at least one entity). The coverage attribute $\acov$ is then defined as the ratio between a solution's number of covered cells and the widest possible coverage in that decision step (belonging to $\sol_0$):
\begin{align}\label{eq:cov}
	\mathrm{cvg}(\sol_n)  \coloneqq&~ \textrm{`number of cells with service in $\sol_n$'} \notag \\
	\acov(\sol_n)  =&~ \frac{\mathrm{cvg}(\sol_n)}{\mathrm{cvg}(\sol_0)}
\end{align}

Unlike $\asig$ and $\asat$, the area of coverage attribute must include all cells covered by the decision maker and its neighboring RSUs. This ensures that new coverage at the edges of the existing network is correctly scored.


\subsubsection{Battery Usage} 
\label{ssub:battery_usage}
To compute the battery usage attribute $\abat$, the car acting as the decision maker must first query its neighbor RSUs for an indicator of their energy expenditure. This request can be sent together with the request for coverage maps that the decision process requires. Vehicles can have different battery capacities, which are in turn affected by the age of the battery, so the indicator is best calculated individually by each vehicle, and not by the decision maker.

In this work, we use a straightforward model where each vehicle has pre-specified standard ($\tau_m$) and maximum ($\tau_M$) times of activity. The indicator decreases linearly once $\tau_m$ is exceeded, and the parked car relinquishes its RSU role forcefully at $\tau_M$: 
\begin{align}\label{eq:battransform}
	\tau_{n,i} \coloneqq&~ \textrm{`active time of entity $\xi_{n,i}$'} \notag \\	
	\mathrm{bat}(\xi_{n,i}) =&~  \left\{ 
		\begin{array}{ll}
			1 & \quad \tau_{n,i} < \tau_m  \\
			1-\frac{\tau_{n,i}-\tau_m}{\tau_M-\tau_m} & \quad \mbox{otherwise}
		\end{array}
	\right.
\end{align}

This design keeps battery concerns invisible to the decision process, at first, and then allows for RSU roles to be reassigned by decision makers before a hard limit on a vehicle's battery resources is reached. 
The resulting battery attribute of a coverage solution is the averaging of all the neighboring roadside units' battery indicators:
\begin{align}\label{eq:bat}
	\xi_n \coloneqq&~ \textrm{`set of entities in solution $\sol_n$'} \notag \\
	|\xi_n| \coloneqq&~ \textrm{cardinality of $\xi_n$} \notag \\
	\xi_{n,i} \coloneqq&~ \textrm{`active entity \textit{i} in solution $\sol_n$'} \notag \\
	\abat(\sol_n) =&~ \frac{\sum_{i\in\xi_n} \mathrm{bat}(\xi_{n,i})}{|\xi_n|}
\end{align}

Coverage solutions that are able to revoke longer-standing RSU roles without affecting the network (e.g., by handing over an older RSU role to a new vehicle in a similar location) will see higher decision scores. We analyze this attribute in detail in Section~\ref{sub:managing_battery_utilization}. 


\subsection{Scoring Algorithm} 
\label{sub:scoring_algorithm}
Pseudocode to score a coverage solution via the Weighted Product Model shown in Section~\ref{sub:ranking_coverage_solutions} is given here.
Algorithm~\ref{alg:scoremetrics}, \textit{ScoreCoverageSolution}, computes the attributes~$a_{j,k}$ and weighs them according to their respective weights~$w_j$, producing the final scoring metric, $\mathrm{score}\left(\sol_k\right)$. These algorithms make use of the decision maker's coverage map ($\SCM_0$), and the coverage maps from its 1-hop and 2-hop neighborhood ($\mathcal{N}_1,\mathcal{N}_2$).
All notation is summarized in Table~\ref{tab:notation}.

\begin{table}[!b]
\centering\small
\caption{Notation Reference}
\label{tab:notation}
\begin{tabular}{@{}p{4.3cm}l@{}}
\toprule
Definition								& Notation					\\ 
\midrule
Matrix indices (row, col.)				& $i,j$ 					\\
Geographic indices (lat., lon.)			& $x,y$						\\
Self-observed coverage map				& $[\SCM] = (\scm_{ij})$		\\
Decision maker's own $\SCM$				& $\SCM_0$		\\
2-hop neighbor $\SCM\mathrm{s}$			& $\mathcal{N}_2 = \{ \SCM_1, \SCM_2, \ldots\}$ \\
Local map of coverage					& $[\LMC] = (\lmc_{ij})$		\\
Local map of saturation					& $[\LMS] = (\lms_{ij})$		\\
Coverage solution	 					& $\sol_k$ \\
Number of covered cells in $\sol_k$	 	& $\mathrm{cov}_{\sol_k}$ \\
$\SCM\mathrm{s}$ of RSUs in $\sol_k$	& $[\SCM_{\sol_k}] = \{\SCM^k_1,\SCM^k_2,\ldots\}$ \\
Battery indicators of RSUs in $\sol_k$	& $[\BAT_{\sol_k}] = (\mathrm{bat}^k_n)$ \\
\bottomrule
\end{tabular}
\end{table}
\begin{algorithm}
\DontPrintSemicolon
\KwData{$\SCM_0$, $[\SCM_{\sol_k}]$, $[\BAT_{\sol_k}]$, $\mathcal{N}_2$, $\mathrm{cov}_{\sol_0}$, $\wsig$, $\wsat$, $\wcov$, $\wbat$}
\KwResult{$\mathrm{score}\left(\sol_k\right)$}
$\triangleright$ $\lmc_{xy}$ (in an $\LMC$) and $\lms_{xy}$ (in an $\LMS$) are initialized to $0$ \;
$\asig \gets \asat \gets \acov \gets \abat \gets 0$ \;
\emph{// create signal and saturation maps}\;
\ForEach{$\SCM_n\{\SCM_{\sol_k},\mathcal{N}_2\}$}{
	\ForEach{$\scm_{n[xy]}\in\SCM_n$}{
	  \lIf{$\scm_{n[xy]} > \lmc_{xy}$}{$\lmc_{xy}\gets{}\scm_{n[xy]}$}
	  \lIf{$\scm_{n[xy]} > 0$}{$\lms_{xy}\gets{}\lms_{xy}+1$}
	}
}
\emph{// average signal strength to get $\asig$} \;
\ForEach{$\lmc_{xy}\in\LMC$}{
	\If{$\lmc_{xy}>0~\mathit{and}~\scm_{xy}\in\SCM_0>0$}{$\asig \gets{} \asig + \lmc_{xy}$}
}
$\asig \gets{} \asig\mathbin{/}\mathrm{cov}_{\SCM_0}$ \;
\emph{// average RSU saturation to get $\asat$} \;
\ForEach{$\lms_{xy}\in\LMS$}{
	\If{$\lms_{xy}>0~\mathit{and}~\scm_{xy}\in\SCM_0>0$}{$\asat \gets{} \asat + \lms_{xy}$}
}
$\asat \gets{} \asat\mathbin{/}\mathrm{cov}_{\SCM_0}$ \;
\emph{// compute $\acov$} \;
$\acov \gets \mathrm{cov}_{\sol_k}\mathbin{/}\mathrm{cov}_{\sol_0}$ \;
\emph{// compute $\abat$} \;
\ForEach{$\mathrm{bat}^k_n \in \BAT_{\sol_k}$}{
	$\abat \gets \abat+\mathrm{bat}^k_n$ \;
}
$\abat \gets \abat\mathbin{/}|\sol_k|$ \;
\emph{// produce a decision score} \;
$\mathrm{score}\left(\sol_k\right) \gets \asig^\wsig \cdot \asat^\wsat \cdot \acov^\wcov \cdot \abat^\wbat$
\caption{ScoreCoverageSolution\label{alg:scoremetrics}}
\end{algorithm}





\section{Evaluation of Self-Organized Networks of Parked Cars} 
\label{sec:evaluation}
We now analyze the performance of the decision process shown in Section~\ref{sec:a_new_approach}. To do so, we run simulations on a \unit[1]{\sqkm} area in the city of Porto, on a custom-designed platform that simulates realistic vehicle mobility~\cite{sumo}, and features real obstruction masks~\cite{postgis}, road topologies~\cite{openstreetmap}, and a communication model that follows empirical signal measurements taken in Porto~\cite{statchannel}. The custom platform and all of its associated data have been made available in~\cite{swift-gissumo}, as part of this research.

The study we present in this section begins by evaluating separate attributes of the decision process in controlled scenarios, to see how each attribute contributes to the resulting RSU network. Then, we create randomized RSU networks and determine best and worst bounds for the self-organized processes, and see how our decision algorithms fare against these theoretical limits. We then look into the decisions' sensitivity to the number of cars on the road and the OBU radio range, and conclude with day-long simulations that integrate realistic models of parking activity and duration.

\subsection{Balancing Roadside Unit Saturation} 
\label{sub:balancing_roadside_unit_saturation}
Signal strength and roadside unit saturation are the two core attributes of our decision process, as they balance one another: stronger mean signal can be reached by assigning new RSU roles in underserved areas, which in turn increases the RSU saturation in the network. 

We begin our study by looking at a decision process with the $\asig$ and $\asat$ attributes alone. We ran sets of 2-hour simulations where the \unit[1]{\sqkm} urban area saw an average of 55 moving vehicles in circulation, entering the city at a rate of 0.5 vehicles per second. The weight of the signal strength, $\wsig$, remained constant at 1.0, while the weight of the saturation attribute, $\wsat$, was adjusted. 

\begin{figure}[t]
	\includegraphics[width=\linewidth]{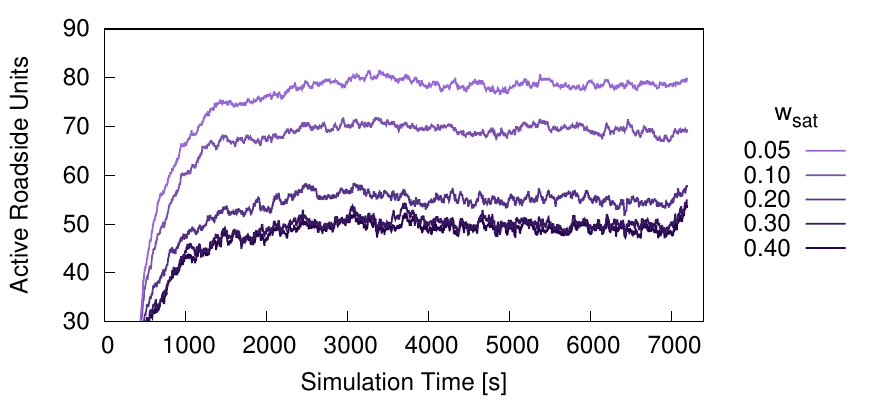}\label{fig:satevolrsu}
	\includegraphics[width=\linewidth]{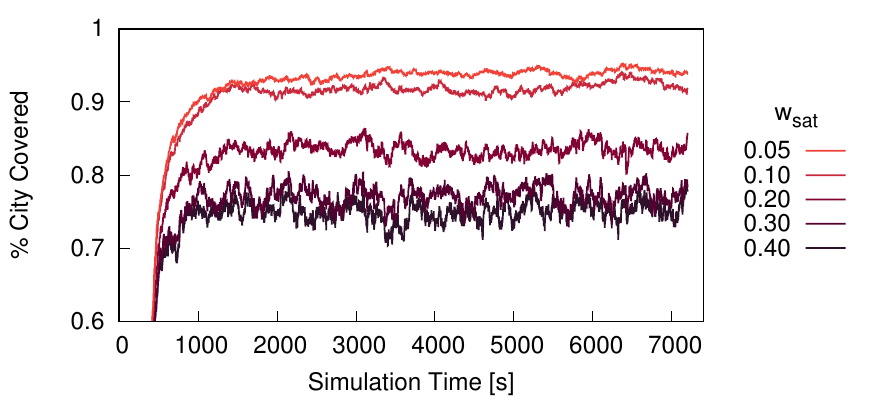}\label{fig:satevolcov}
	\caption{Parked cars with active RSU roles~(top) and percentage of the urban area where RSU service is made available~(bottom), through the course of 2-hour simulations, as the weight controlling the RSU saturation attribute is varied. Data sampled in 1-second intervals. }
	\label{fig:satevolution}
\end{figure}

The evolution of the number of active RSUs and the percentage of the city covered by those RSUs is plotted in Figure~\ref{fig:satevolution}, with samples taken in 1-second intervals. Beginning with a network devoid of RSUs, the transient state of the network lasted for approximately 1060 seconds ($\approx$\unit[17]{min.}) from the moment the first vehicle parked until the city's coverage area stabilized. Both figures show that the network of parked cars enters a steady state, and these decision mechanisms keep the network stable in terms of RSUs and coverage. The same behavior was seen in the mean signal strength and mean RSU saturation (not shown).

We present a steady state analysis of the network in Table~\ref{tab:satvary}, as the weight $\wsat$ that controls the RSU saturation attribute is increased. Both attributes behave as expected: higher weights on $\wsat$ push towards fewer active RSUs and, consequently, a lower mean signal strength (as the signal strength attribute, $\asig$, is de-prioritized in favor of $\asat$). The metrics' standard deviations become tighter depending on which attribute is leading the decision process, which is desirable.

\begin{table}[!b]
	\small
	\centering
	\caption{Network Steady State Varying a Saturation Attribute}
	\label{tab:satvary}
	\renewcommand{\arraystretch}{1.5}
	\setlength\tabcolsep{5pt}
\begin{tabular}{@{}cccccc@{}}
\toprule
$\wsat$ & \tabmedstack{Signal Distribution\\\signalDistKey} & \tabmedstack{City\\Cov.} & \tabmedstack{Signal\\Str.} & \tabmedstack{RSU\\Sat.} & \tabmedstack{Active\\RSUs} \\ \midrule
0.05 & \raisebox{-.20\height}{ \includegraphics[width=90pt,height=8pt]{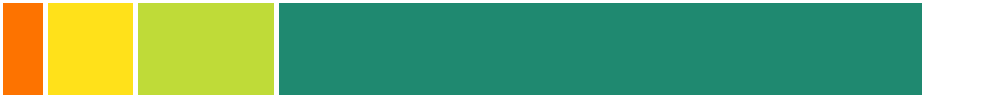} } & 94\% & \tabshortstack{4.51\\$\scriptscriptstyle\pm0.86$} & \tabshortstack{2.62\\$\scriptscriptstyle\pm1.11$} & 79 \\
0.10 & \raisebox{-.20\height}{ \includegraphics[width=90pt,height=8pt]{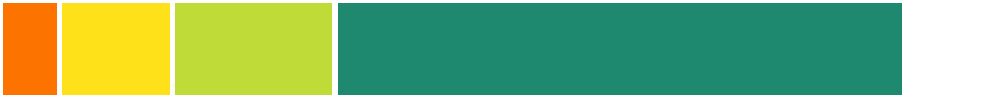} } & 92\% & \tabshortstack{4.38\\$\scriptscriptstyle\pm0.93$} & \tabshortstack{2.21\\$\scriptscriptstyle\pm0.95$} & 69 \\
0.20 & \raisebox{-.20\height}{ \includegraphics[width=90pt,height=8pt]{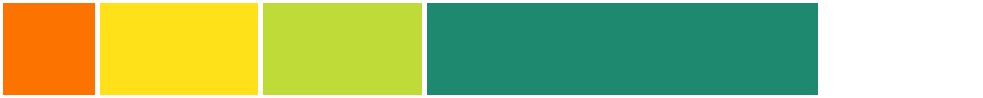} } & 83\% & \tabshortstack{4.05\\$\scriptscriptstyle\pm1.07$} & \tabshortstack{1.64\\$\scriptscriptstyle\pm0.72$} & 55 \\
0.30 & \raisebox{-.20\height}{ \includegraphics[width=90pt,height=8pt]{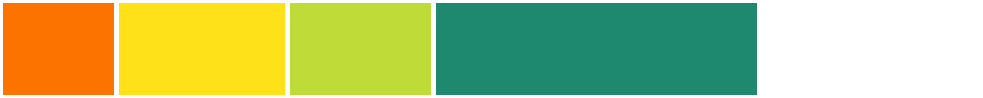} } & 77\% & \tabshortstack{3.90\\$\scriptscriptstyle\pm1.12$} & \tabshortstack{1.48\\$\scriptscriptstyle\pm0.65$} & 50 \\
0.40 & \raisebox{-.20\height}{ \includegraphics[width=90pt,height=8pt]{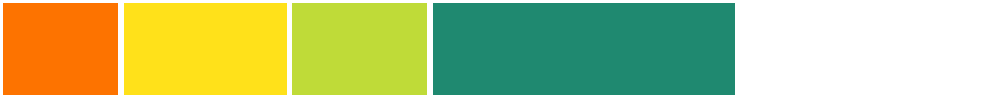} } & 75\% & \tabshortstack{3.86\\$\scriptscriptstyle\pm1.13$} & \tabshortstack{1.45\\$\scriptscriptstyle\pm0.63$} & 49 \\
\bottomrule
\multicolumn{6}{r}{\scriptsize $\wsig=1.0$}
\end{tabular}
\end{table}


We continued to push $\wsat$ beyond $0.4$, but no meaningful changes were observed in the network. The best (lowest) RSU saturation that was observed was of $1.4$ RSUs per cell. This self-organized process is unable to push for lower saturation levels (ideally, towards the minimum of 1.0, a single RSU per cell), for the understandable reason that it requires communication between RSUs in order to function. For a pair of RSUs to communicate with one another, their coverages must necessarily overlap -- therefore, RSU saturation in that overlap must exceed $1.0$.

Increasing $\wsat$ pushes the saturation of the network down, decreasing the number of RSUs, while also increasing the efficiency in terms of urban area covered per RSU. The \unit[1]{\sqkm} urban area in Porto where we run our simulations has 650 usable cells (i.e., where vehicles can move through), which corresponds to a \unit[468\,000]{\sqm} (\unit[0.47]{\sqkm}) area that the parked cars can provide coverage to. At the lowest $\wsat$, the average area covered per RSU was of \unit[5\,600]{\sqm}, while at the highest $\wsat$, each RSU was able to cover \unit[7\,200]{\sqm}, a 28\% increase. We observed a strong inverse correlation between RSU saturation and urban area covered per RSU (at a linear correlation coefficient of $-0.995$).


\subsection{Expanding Network Coverage} 
\label{sub:expanding_network_coverage}
The previous analysis showed that restricting the number of RSUs in the city causes the undesired side effect of reducing the citywide coverage that the RSU network is able to provide. While this is expected (coverage and RSU count are inherently linked), a scoring algorithm must be able to optimize the network towards providing a more widespread coverage, as well. We now see how the inclusion of a coverage attribute into the decision process helps achieve this goal.

\begin{figure}[t]
	\centering
	\includegraphics[width=\linewidth]{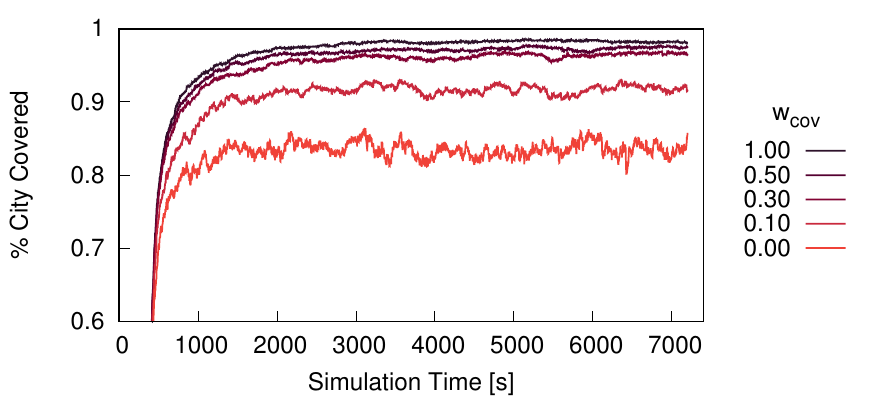}
	\caption{Percentage of the urban area covered by the self-organized RSU network, through the course of 2-hour simulations, as the weight controlling the coverage attribute is varied. Data sampled in 1-second intervals.}
	\label{fig:covevolution}
\end{figure}

We repeated the 2-hour simulation sets, fixing $\wsat=0.2$ to serve as a baseline from the previous analysis, while increasing the weight $\wcov$ of the coverage attribute. Figure~\ref{fig:covevolution} shows how the percentage of the city that is covered evolves throughout the simulation, for the different attribute weights. The data show that the coverage attribute pushes the RSU network's total coverage to high levels, and improves the stability of that coverage. The duration of the transient state does not appear to change through different attribute sets, as well.

We provide an analysis of the network's steady state in Table~\ref{tab:covvary}. The data show that higher $\wcov$ values lead to a more widespread coverage of the network with fewer RSUs than in our initial analysis. To push towards better coverage, the decision process does recruit additional RSUs, increasing both signal strength and RSU saturation as a result. However, it does so at a more optimal distribution of RSUs (more area covered per RSU) than before. To illustrate this fact, Table~\ref{tab:covcompare} compares two parameter sets that recruit similar numbers of RSUs and achieve similar mean signal strength in the city, the first set with $\asig$ and $\asat$ attributes only, and the second with the $\acov$ attribute included in the scoring algorithm. This comparison shows that the inclusion of $\acov$ increased citywide coverage by 5\% and reduced RSU saturation by 8\%, at no increase to the number of active RSUs or cost to the mean signal strength that is being provided.

With this analysis, we showed that a coverage attribute improves the self-organization of the parked cars, ensuring not only a higher citywide coverage but also a better distribution of RSU roles, with lower RSU saturation and higher area covered per RSU.



\subsection{Managing Battery Utilization} 
\label{sub:managing_battery_utilization}
The fourth and final attribute of our decision scoring algorithm concerns the time a parked car spends with its DSRC electronics activated, performing RSU roles, and drawing power from the car's battery. A vehicle's battery, when its engine is off, is a finite resource that must be managed correctly. Not only must the car's energy budget be shared with electrical systems that are active when the car parks (e.g., alarm systems, and keyless entry systems), but failing to do so has the serious consequence of leaving the driver stranded, should the battery be too drained for engine re-ignition.

\begin{table}[!b]
	\small
	\centering
	\caption{Network Steady State Varying a Coverage Attribute}
	\label{tab:covvary}
	\renewcommand{\arraystretch}{1.5}
	\setlength\tabcolsep{5pt}
\begin{tabular}{@{}cccccc@{}}
\toprule
$\wcov$ & \tabmedstack{Signal Distribution\\\signalDistKey} & \tabmedstack{City\\Cov.} & \tabmedstack{Signal\\Str.} & \tabmedstack{RSU\\Sat.} & \tabmedstack{Active\\RSUs} \\ \midrule
0.00 & \raisebox{-.20\height}{ \includegraphics[width=90pt,height=8pt]{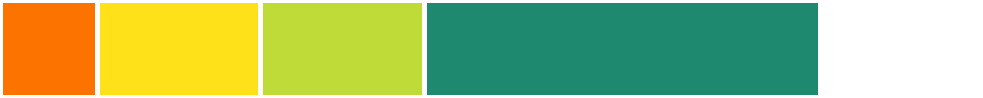} } & 83\% & \tabshortstack{4.05\\$\scriptscriptstyle\pm1.07$} & \tabshortstack{1.64\\$\scriptscriptstyle\pm0.73$} & 55 \\
0.10 & \raisebox{-.20\height}{ \includegraphics[width=90pt,height=8pt]{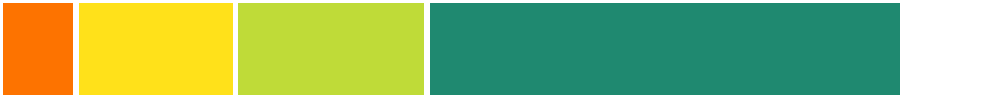} } & 92\% & \tabshortstack{4.18\\$\scriptscriptstyle\pm1.00$} & \tabshortstack{1.77\\$\scriptscriptstyle\pm0.77$} & 60 \\
0.30 & \raisebox{-.20\height}{ \includegraphics[width=90pt,height=8pt]{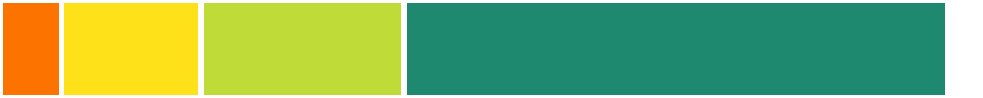} } & 96\% & \tabshortstack{4.30\\$\scriptscriptstyle\pm0.94$} & \tabshortstack{1.94\\$\scriptscriptstyle\pm0.82$} & 65 \\
0.50 & \raisebox{-.20\height}{ \includegraphics[width=90pt,height=8pt]{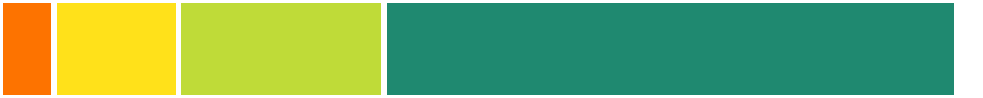} } & 97\% & \tabshortstack{4.36\\$\scriptscriptstyle\pm0.90$} & \tabshortstack{2.03\\$\scriptscriptstyle\pm0.84$} & 68 \\
1.00 & \raisebox{-.20\height}{ \includegraphics[width=90pt,height=8pt]{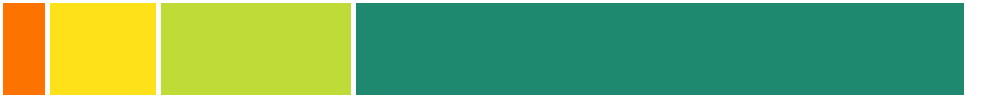} } & 98\% & \tabshortstack{4.43\\$\scriptscriptstyle\pm0.87$} & \tabshortstack{2.19\\$\scriptscriptstyle\pm0.85$} & 72 \\
\bottomrule
\multicolumn{6}{r}{\scriptsize $\wsig=1.0$, $\wsat=0.2$}
\end{tabular}
\end{table}

\begin{table}[!b]
	\small
	\centering
	\caption{Parameter Set Comparison in Steady State}
	\label{tab:covcompare}
	\renewcommand{\arraystretch}{1.5}
	\setlength\tabcolsep{5pt}
\begin{tabular}{@{}ccccccc@{}}
\toprule
\tabshortcstack{$\wsat$\\$\wcov$} & \tabmedstack{Signal Distribution\\\signalDistKey} & \tabmedstack{City\\Cov.} & \tabmedstack{Signal\\Str.} & \tabmedstack{RSU\\Sat.} & \tabmedstack{Active\\RSUs} \\ \midrule
\tabshortstack{\footnotesize 0.10\\\footnotesize 0.00} & \raisebox{-.23\height}{ \includegraphics[width=90pt,height=8pt]{D0_sigSatHorizCovDist/wsat010} }  & 92\% & \tabshortstack{4.38\\$\scriptscriptstyle\pm0.93$} & \tabshortstack{2.21\\$\scriptscriptstyle\pm0.95$} & 69 \\
\tabshortstack{\footnotesize 0.20\\\footnotesize 0.50} & \raisebox{-.23\height}{ \includegraphics[width=90pt,height=8pt]{E0_wideCovHorizCovDist/wcov050} } & 97\% & \tabshortstack{4.36\\$\scriptscriptstyle\pm0.90$} & \tabshortstack{2.03\\$\scriptscriptstyle\pm0.84$} & 68 \\
\bottomrule
\multicolumn{6}{r}{\scriptsize $\wsig=1.0$}
\end{tabular}
\end{table}

By including the battery utilization of active RSUs in the scoring algorithm, we aim to drive the decision makers towards coverage solutions that replace aging RSUs with newly parked cars, while at the same time attempting to preserve the existing structure of the RSU network. 

To study the efficacy of this attribute, we begin by setting a hard threshold of 1~hour on the maximum time a parked vehicle can hold RSU roles for. Active parked cars that exceed this threshold are forcefully removed from the RSU network. We then set the $\tau$ parameters in Equation~\eqref{eq:battransform} to begin penalizing RSUs once their active time exceeds 30~minutes ($\tau_m=1800,\tau_M=3600$). With this configuration, RSU battery life does not factor into the decision process until 30~minutes of activity have elapsed, allowing for decisions to focus exclusively on the quality of the network as long while the parked cars' activity times are within expected values.
Empirical data provided in~\cite{tits2017} showed that an OBU operating for \unit[$\approx$6.5]{hours} consumes approximately $4.2\%$ of a typical car battery. We opt for a more conservative limit in this evaluation: 30~minutes to 1~hour with an active OBU would result in a battery usage of $0.32\%-0.64\%$.

The figures in Table~\ref{tab:batvary} show the distribution of the lifetime of RSU roles in the network during 8-hour simulations, together with the network's metrics after the transient state, as we increase the weight $\wbat$ of the battery attribute.
\begin{table}[!b]
	\small
	\centering
	\caption{Roadside Unit Lifetime Analysis}
	\label{tab:batvary}
	\renewcommand{\arraystretch}{1.5}
	\setlength\tabcolsep{5pt}
\begin{tabular}{@{}cccccc@{}}
\toprule
$\wbat$ & \tabmedstack{RSU Lifetime\\\histKey} & \tabmedstack{City\\Cov.} & \tabmedstack{Signal\\Str.} & \tabmedstack{RSU\\Sat.} & \tabmedstack{Act.\\RSUs} \\ \midrule
0.00 & \raisebox{-.35\height}{ \includegraphics[width=80pt,height=21pt]{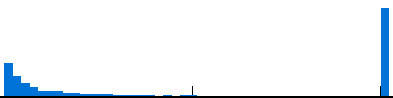} } & 95\% & \tabshortcstack{4.24\\$\scriptscriptstyle\pm0.98$} & \tabshortcstack{2.09\\$\scriptscriptstyle\pm0.85$} & 68 \\
0.30 & \raisebox{-.35\height}{ \includegraphics[width=80pt,height=21pt]{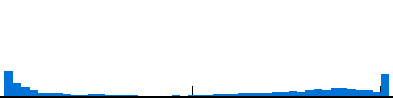} } & 96\% & \tabshortcstack{4.29\\$\scriptscriptstyle\pm0.97$} & \tabshortcstack{2.13\\$\scriptscriptstyle\pm0.84$} & 70 \\
0.50 & \raisebox{-.35\height}{ \includegraphics[width=80pt,height=21pt]{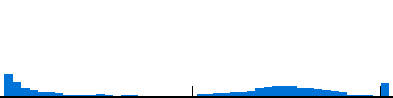} } & 96\% & \tabshortcstack{4.28\\$\scriptscriptstyle\pm0.95$} & \tabshortcstack{2.12\\$\scriptscriptstyle\pm0.86$} & 70 \\
1.00 & \raisebox{-.35\height}{ \includegraphics[width=80pt,height=21pt]{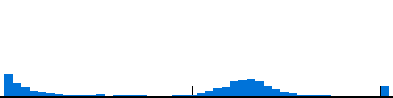} } & 95\% & \tabshortcstack{4.27\\$\scriptscriptstyle\pm0.96$} & \tabshortcstack{2.10\\$\scriptscriptstyle\pm0.85$} & 69 \\
1.50 & \raisebox{-.35\height}{ \includegraphics[width=80pt,height=21pt]{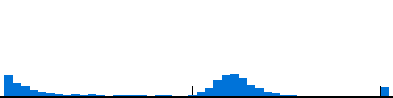} } & 95\% & \tabshortcstack{4.26\\$\scriptscriptstyle\pm0.97$} & \tabshortcstack{2.10\\$\scriptscriptstyle\pm0.84$} & 69 \\
\bottomrule
\multicolumn{6}{l}{\scriptsize Lifetime histograms span 3600 seconds. $\wsig=1.0$, $\wsat=0.2$, $\wcov=0.3$}
\end{tabular}
\end{table}

When $\abat$ is not in play ($\wbat=0$), we see most RSUs being disabled at our designated 1-hour threshold. A noticeable number of RSU roles are also revoked shortly after they are assigned, which can be attributed to the initial set of roles being rotated out as newly parked cars in more optimal locations become available. As $\wbat$ is increased, the RSUs that were hitting the 1-hour threshold begin to be replaced instead by decisions from decision makers in the city, with higher weights leading to earlier role revocation, pushing the lifetime of RSUs closer towards $\tau_m$.

The steady-state analysis of various city metrics in Table~\ref{tab:batvary} indicates that a well-designed attribute can rotate RSU roles efficiently, with a negligible impact to the existing self-organized RSU network. Mean signal strength, RSU saturation, city coverage, and number of active RSUs all deviated less than 3\% as $\wbat$ was increased, relative to the baseline where the battery attribute was not included. This important result shows the efficacy of our proposed algorithm in the management of a parked car's battery utilization as it takes on RSU tasks.

Finally, we note that a small percentage of RSU roles still exceed the 1-hour threshold and have to be forcefully revoked, even at the highest $\wbat$ weights we evaluated. We hypothesize that these are RSUs in critical locations where not enough new parking events occur for a decision maker to be able to find an adequate replacement. Increasing the time allotted for an RSU role to be replaced ($\tau_M-\tau_m$) may mitigate this behavior.


\subsection{Decision Performance} 
\label{sub:decision_performance}
\begin{figure}[t]
	\centering
	\includegraphics[width=\linewidth]{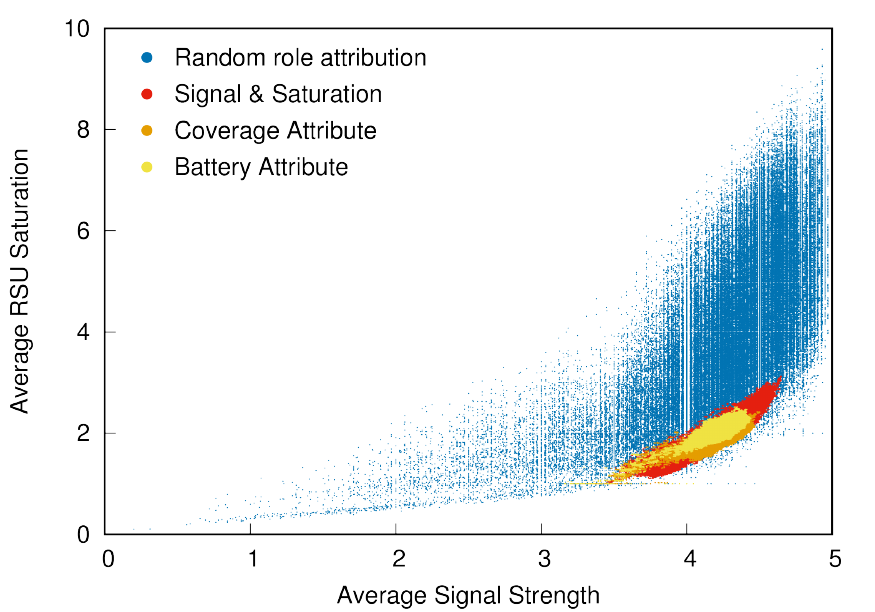}
	\caption{Signal strength and RSU saturation metrics sampled in 1-second intervals from the simulations in Sections~\ref{sub:balancing_roadside_unit_saturation}, \ref{sub:expanding_network_coverage} and~\ref{sub:managing_battery_utilization}.}
	\label{fig:overlay_all}
\end{figure}

An essential measure of the quality of a self-organized network of parked cars is its ratio of signal strength to RSU saturation. For a certain mean signal quality that is achieved in a given area, there is an optimal and a sub-optimal number and placement of RSUs that may lead to it. 
We study this ratio in our simulation setup by filling the network with parked vehicles, and assigning RSU roles to those vehicles in a random manner. For each random attribution of roles, we then measure the resulting signal strength and RSU saturation in the network. The data in Figure~\ref{fig:overlay_all} show 500.000 random assignments, on top of which we overlay data of the citywide metrics that were collected during the simulations presented in the previous three sections.

Relatively clean upper and lower bounds can be observed from the sets of random attributions. These bounds are an important finding: for self-organized networks of cars, they reveal that the ratio between the number of recruited cars and the strength of the coverage being provided will be bounded. Here, the lower bound represents the most optimal placement of RSUs for each observation of signal strength, while the opposite is true for the upper bound. This gives us an important frame of reference to analyze the performance of any decision making process.
The data show that an increasing effort is required to push the network to higher signal strengths. For example, improving mean signal from $2.0$ to $4.0$ is possible with, on average, one extra RSU, however the same effort is required to move the mean signal from $4.0$ to $4.5$, and more than double that effort will be needed to go from $4.5$ to $4.9$. This assumes an optimal assignment of RSU roles, which might not always be possible.


The overlaid data from the decision mechanisms demonstrate that these operate very close to the lower signal-to-saturation bound revealed by the random role assignment data. The scoring algorithm that included a coverage attribute, in Section~\ref{sub:expanding_network_coverage}, operates closer to this lower bound than the initial 2-attribute algorithm, which matches our earlier observation that this attribute may lead to a more optimal distribution of RSU roles. The introduction of an RSU lifetime threshold and the inclusion of a battery attribute, in Section~\ref{sub:managing_battery_utilization}, sees the data samples pushed away from this ideal lower bound to a small degree, which is an expected consequence from the forceful revocation of RSU roles from parked cars that have been active for some time. In all three cases, regardless, the decision algorithms place the resulting RSU network in a very optimal region.

These important results show that our decision mechanisms are operating near the best observed efficiency for a network comprised of parked cars, achieving various levels of signal strength at the smallest possible RSU saturation and, consequently, the lowest number of RSUs at their most optimal arrangement.


\subsection{Sensitivity to Vehicle Density} 
\label{sub:sensitivity_to_vehicle_density}
The simulation studies presented so far saw an average number of 55 moving vehicles per \sqkm~on the road, which matches a medium-low density scenario. We now analyze identical scenarios where we reduce the number of moving vehicles, to find out how the decision mechanisms respond to the lower vehicle density. This density affects primarily the time parked cars take to build their coverage maps, and the frequency of parking events. 

\begin{figure}[t]
	\centering
	\includegraphics[width=\linewidth]{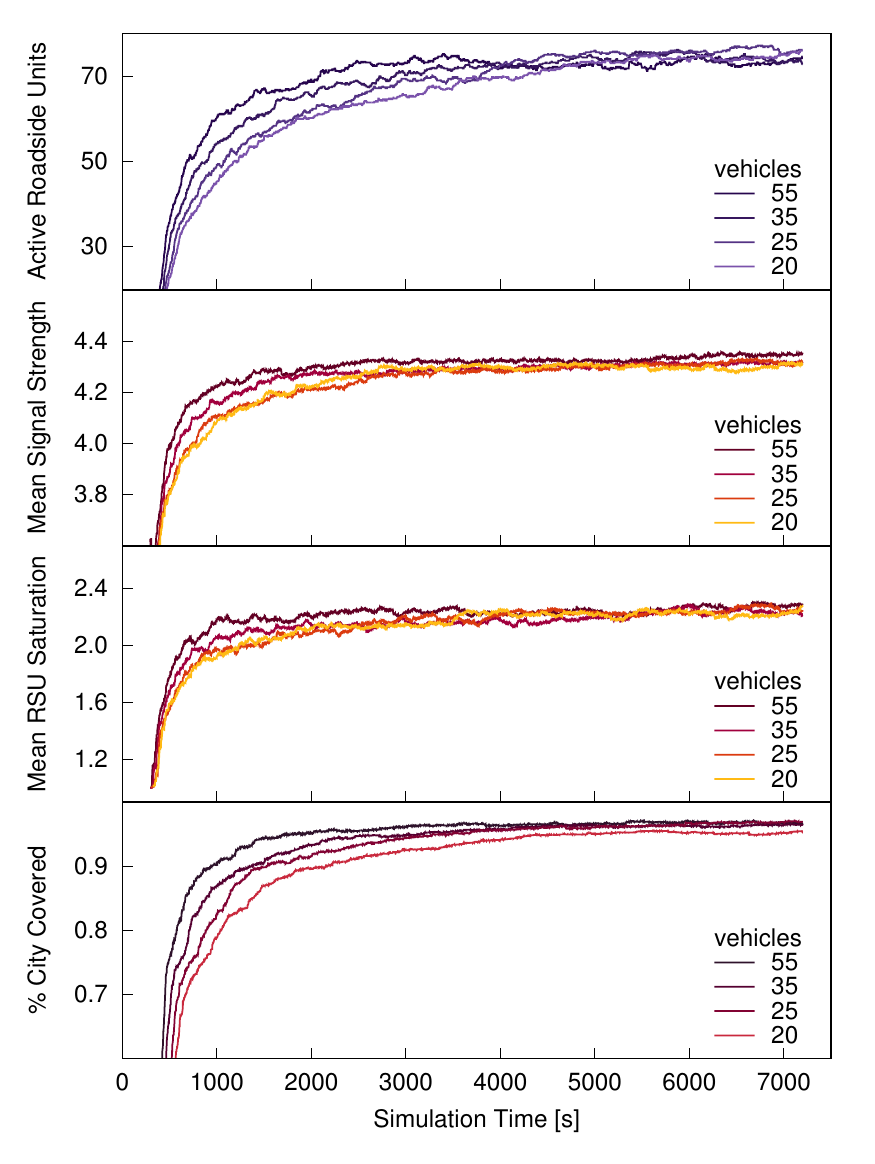}
	\caption{Evolution of various network metrics throughout 2-hour simulations, as the density of vehicles per \sqkm~is altered. Scoring algorithm weights set to $\wsig=1$, $\wsat=0.2$, $\wcov=0.3$, and $\wbat=0$. Data sampled in 1-second intervals.}
	\label{fig:periodTetra}
\end{figure}

The evolution of the number of active RSUs, mean signal strength, mean RSU saturation, and total city coverage can be seen in Figure~\ref{fig:periodTetra}. Four densities of vehicles were considered: 55, 35, 25, and 20 vehicles per~\sqkm, which range from medium- through low-density scenarios. 

The data in these figures show that only the duration of the network's transient state is affected by the decrease in the number of road vehicles, with all four metrics eventually converging to a similar steady state. This indicates that the decision mechanisms are robust and should operate equally well in the face of varying vehicle densities, as is expected from real life scenarios. A small 2\% decrease can be seen in the steady state citywide coverage, which could indicate that the lowest-density scenarios have insufficient new parking events for the algorithm to completely optimize the network.

\begin{figure}[t]
	\centering
	\includegraphics[width=0.49\linewidth]{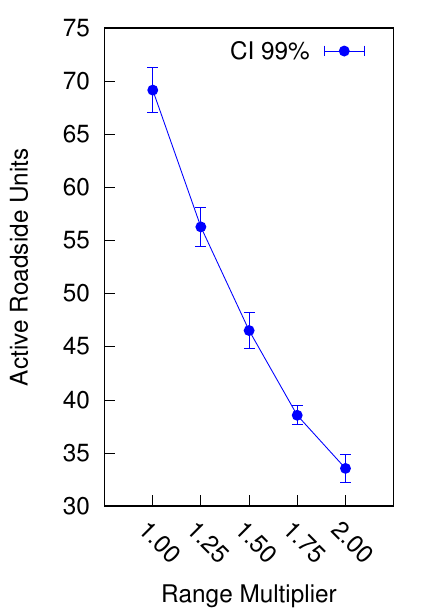}
	\includegraphics[width=0.49\linewidth]{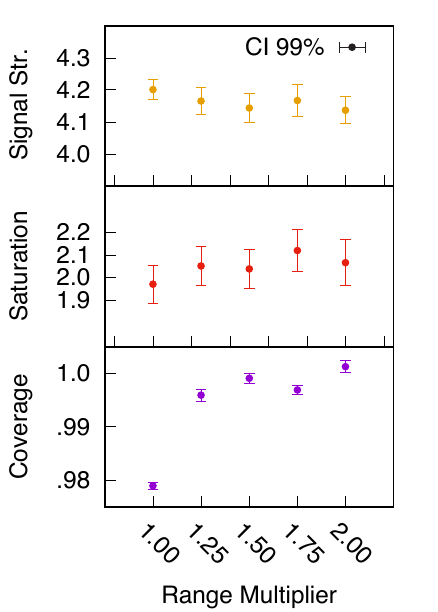}
	\caption{Steady state averages of various network metrics of 2-hour simulations, as a multiplier is applied to the radio range of vehicles and parked cars. Scoring algorithm weights set to $\wsig=1$, $\wsat=0.3$, $\wcov=0.5$, and $\wbat=0$. Confidence intervals stated in the 99\% level.}
	\label{fig:rangeMultiActRsu}
\end{figure}


\subsection{Sensitivity to Wireless Radio Range} 
\label{sub:sensitivity_to_wireless_radio_range}
We now provide an analysis of the decision mechanisms' sensitivity to the radio range of the OBUs used in the vehicles. Figure~\ref{fig:rangeMultiActRsu} shows various network metrics in the network's steady state as a multiplier is applied to our propagation models, scaling the effective range of the vehicles' DSRC radios. 
The data indicate that the number of RSU roles assigned by the self-organizing mechanisms decreases almost linearly with the corresponding increase in radio range: doubling the radio range halves the number of RSUs, from $\approx69.1$ to $\approx33.6$. This is an interesting result due to the fact that the total area that a vehicle's radio can reach should increase quadratically with the radio range ($\mathrm{area}=\pi r^2$), assuming perfect conditions for propagation. In an urban environment, however, vehicles's signals are constrained by the road layout, in effect being tunneled by surrounding buildings. We hypothesize this to be the reason behind the near-linear relation that is observed, instead of the quadratic one that one might expect.

Besides the number of active RSUs, both signal strength and RSU saturation remain relatively constant, despite the change in radio range, indicating that the algorithms are resilient to this network aspect. In terms of total coverage being provided, the range increase also grew this metric by 2\%, to a near-complete coverage level. 


\subsection{Realistic Parking} 
\label{sub:realistic_parking}
\begin{figure}[t]
	\centering
	\includegraphics[width=\linewidth]{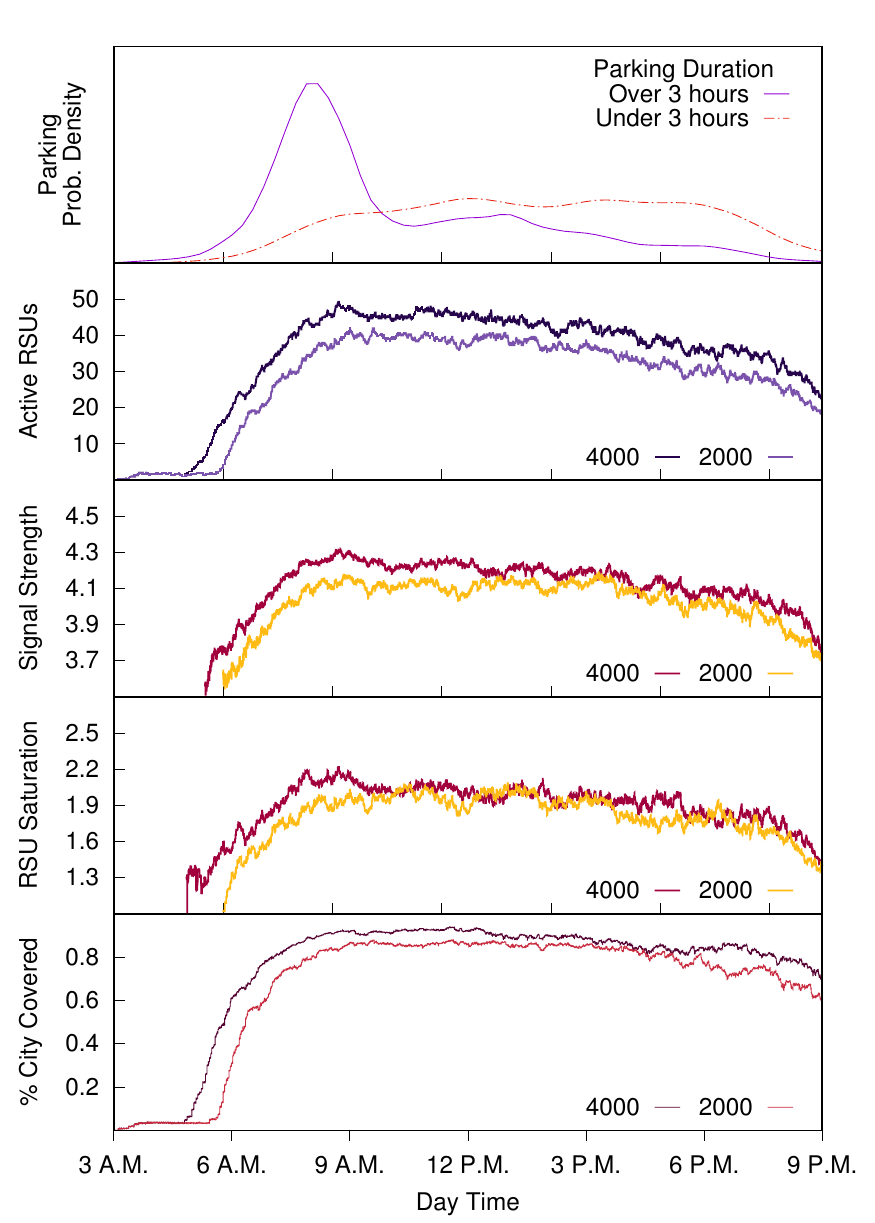}
	\caption{Evolution of various network metrics throughout whole-day simulations with realistic parking behavior. The first plot shows a simplified view of how parking events are distributed, with long-term and short-term parking behaviors split up. The remaining plots compare two densities of parking events: 4000 and 2000 vehicles parked (over 24 hours). Scoring algorithm weights set to $\wsig=1$, $\wsat=0.2$, $\wcov=0.3$, and $\wbat=0$.}
	\label{fig:real5up}
\end{figure}

We conclude our simulation work with a look at the self-organizing mechanisms operating under realistic parking conditions. In order to do so, we integrated the realistic parking models presented in~\cite{smartvehicles16}, which include a distribution of vehicle arrivals (and subsequent parking) throughout the day, and a comprehensive hour-by-hour stochastic model of the time these same vehicles remain parked. The data from whole-day simulations performed with accurate parking models is presented in~Figure~\ref{fig:real5up}. Parking events at two densities (4000 and 2000 vehicles parked over 24 hours) are distributed throughout the day following the models from~\cite{smartvehicles16}. 

In realistic scenarios, decision mechanisms face two new issues. The first is a possible shortage of newly parked cars, which constrains the assignment of RSU roles to vehicles that may not always be at locations optimal for an RSU network. The second concerns the vehicles's parking durations: evidently, cars parking for only a short time make for poor RSUs, and this parking duration is not known beforehand; but a more pernicious effect occurs when a car parking for a short time makes the decision to disable a nearby car that is parked for a longer time. Without knowledge of the time each vehicle will spend parked, or a means to recruit previously-disabled RSUs, this side effect is difficult to mitigate.

Despite these issues, the data in~Figure~\ref{fig:real5up} show that, under realistic parking conditions, a self-organized network of parked cars is able to provide and maintain widespread coverage of the urban area throughout the day, at adequate levels of coverage strength, while assigning roles to only a fraction of the vehicles that park throughout the day.

\begin{figure*}[!t]
	\centering
	\includegraphics[width=1.0\linewidth]{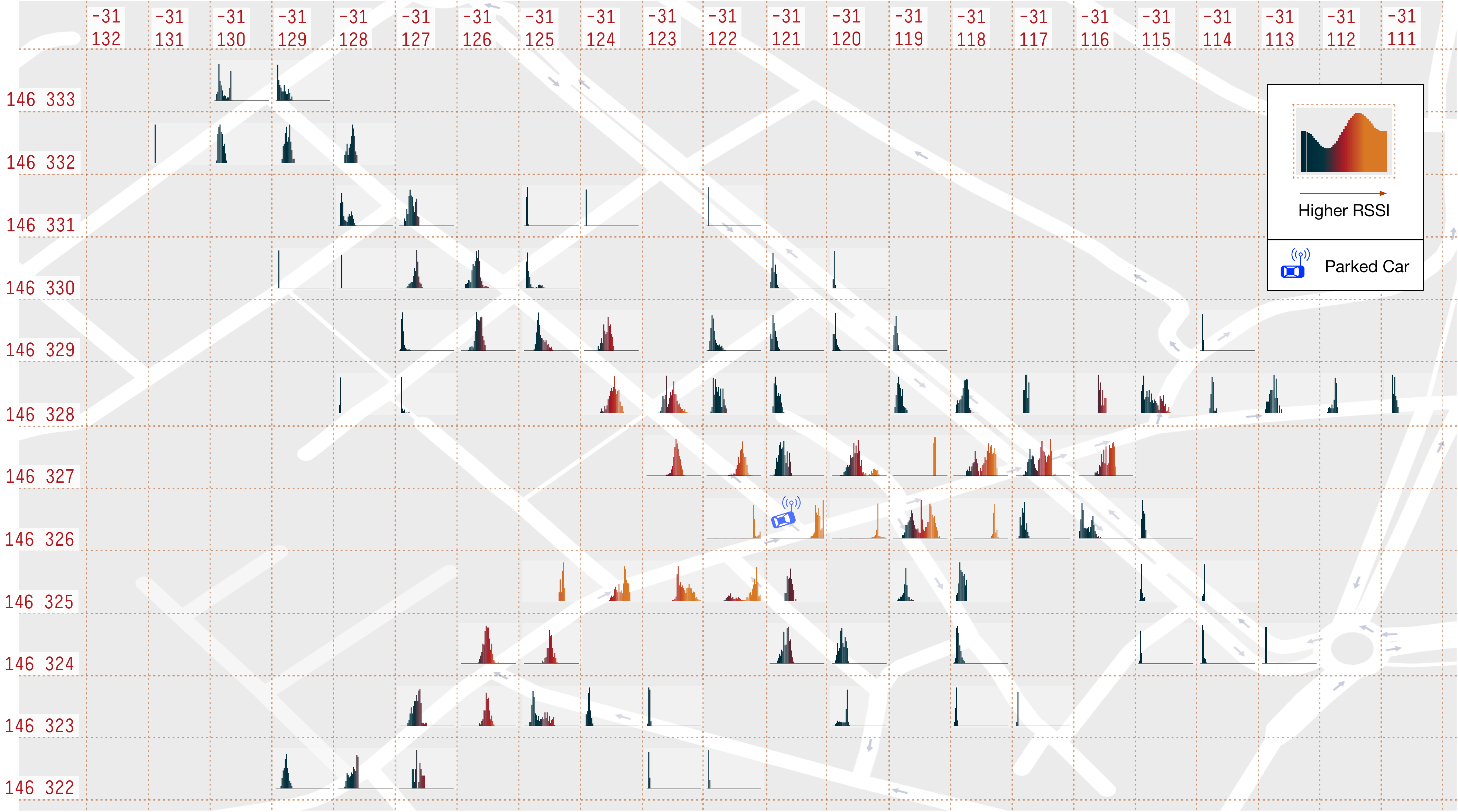}
	\caption{Histograms of signal strength measurements made by a parked car (in blue) of beacons received from other vehicles that were traveling through neighboring cells. Each histogram shows the distribution of the RSSI of all beacons overheard from the cell it is plotted in. The parked car was positioned near an intersection on cell (-31\,121, 146\,326).}
	\label{fig:histograms}
\end{figure*}

Practical applications for parked cars may benefit from attempting to predict the parking duration for each vehicle, prioritizing those more likely to remain parked in RSU role assignment. As an example, the models in~\cite{smartvehicles16} indicate that long-term parking occurs primarily in the early morning hours, so vehicles that park in these hours may be more suitable candidates for RSU roles.





\section{Inferring Coverage Maps from Received Beacons} 
\label{sec:empirical}
In this section we present an empirical study that replicates the learning process a newly parked car goes through in order to determine its own coverage map.
A primary concern of any self-organizing approach that relies on a parked car's ability to determine its own coverage map so that informed decisions can be made, is the reliability of the process that creates the car's coverage map. Radio signals can vary greatly in an urban environment, due to the significant number of obstructions that cause the signal to fade. Wave reflections and corner diffraction also contribute to the received signal in an urban environment.


We performed an experimental study of the process of creating a coverage map by a parked vehicle. For this study, we parked a DSRC-enabled vehicle and instructed it to collect beacons being broadcast from neighboring vehicles. These beacons contain the GPS coordinates of the transmitting vehicles, and the parked car can determine the signal strength (RSSI) of each received beacon. We then drove other vehicles equipped with OBUs in the neighboring roads of the parked car.
This experiment was located in the city of Aveiro, Portugal, in an urban area with a variety of small residences, low-rise apartments, and commercial buildings. 


From the beacon data that was recorded by the parked car, the car then grouped measurements by their geographic cell, and computed the average signal strength and standard deviation of the measurements belonging to each cell. We then plotted histograms of signal strength to better understand its distribution. These can be seen in~Figure~\ref{fig:histograms}, overlaid on top of a city map.


These histograms show that vehicle-to-vehicle signal strength in an urban vehicular network is well-behaved over short distances. This suggests that a self-organizing approach can depend on coverage maps that were learned by parked cars (by observing beacons), and that the signal strength values estimated for each cell will be representative of the strength of coverage that can be provided to that cell by the listening vehicle.
On most cells, the spread of the signal strength around the mean is relatively small, particularly if one considers how dynamic the environment can be. Our OBUs report RSSI in a range of 1 to 50, and of the 75 cells from which beacons were received, 88\% of the cells showed a standard deviation under 5.0.

The nature of a cell-based coverage maps does imply that a level of abstraction is applied to the city map and, consequently, there may be situations where a deep fade occurs within a cell, and the average received power from beacons fails to reflect the coverage quality on that cell. In these situations, a bimodal distribution can often be seen. 


\section{Related Work}
\label{sec:relatedwork}
Monitoring and collecting data on vehicular traffic, road congestion, and trip patterns are natural uses for DSRC technology, where information can be pulled directly from DSRC-equipped vehicles, obviating the need for more complex approaches. Research in this field has suggested the use of vehicles~\cite{alam2011}, buses~\cite{lan2008}, and infrastructure-supported VANETS~\cite{felice2014} as viable candidates for gathering such data. 
The location and tracking of parking space availability is another possible use of DSRC technology, and proposals for detecting and distributing this type of data can be found in~\cite{caliskan2006,panayappan2007}. For public transportation, the work in~\cite{wang2002} proposes the use of DSRC to capture real-time information on the location of city buses, as well as the traffic conditions that surround the buses' routes.  Vehicular networks have also been shown to be effective collectors and disseminators of sensor data~\cite{lee2006}, which are crucial for realizing the vision for a Smart City. In certain conditions, the vehicles themselves can also become sensors~\cite{bazzi2015}, performing street imaging, temperature and weather monitoring, or license plate recognition. A common theme to this body of work is that either infrastructure is required to gather and distribute the required data, or opportunistic, delay-tolerant mechanisms are resorted to in order to improve the probability of data collection.

The concept of leveraging parked cars as an alternative to expensive deployments of roadside units was first introduced in~\cite{pvavanet}, as a way to boost node density in a sparse vehicular network. Interesting works such as~\cite{dressler2014} and~\cite{dressler2012} further contributed to this idea, suggesting the use of parked cars to overcome communication issues due to corner obstructions, and as RSU assistants to content caching and downloading. Later works, in~\cite{vtc2015,tits2017}, proposed a broader use for these entities, allowing them to self-organize to form a more expansive and versatile vehicular support network. 

Our work brings new mechanisms for self-organizing parked cars that improve on existing approaches by being able to further optimize the resulting support network, acting both during the initial formation of the network and its steady state, and incorporating battery usage information to limit the power drawn from each active car. This work also provides the first comprehensive evaluation of such networks in scenarios with realistic models of parking frequency and duration, and an experimental verification of key concepts.

\section{Conclusion} 
\label{sec:conclusion}
A novel approach has been provided to enable parked cars to self-organize and form widespread vehicular support networks in the urban area, which can then be used to monitor vehicular traffic, public transportation, and available parking. 
Through newly introduced mechanisms and a multi-criteria decision process, the vehicular support network that is created can be continuously optimized through time, designed to target specific metrics of signal strength, roadside unit saturation, and area covered, while rotating RSU roles between vehicles to manage battery utilization. 
An extensive simulation study, coupled with an experimental verification of key concepts, provides the first strong evaluation of a self-organizing approach for parked cars. It reveals, for the first time, how such a network forms and evolves, what quality of coverage is possible and how it is distributed, what balance is achievable between recruited vehicles and network strength, and validates the proposed mechanisms against simulations that integrate realistic models of parking behavior.

\begin{IEEEbiography}
[{\includegraphics[width=1in,height=1.25in,clip,keepaspectratio]{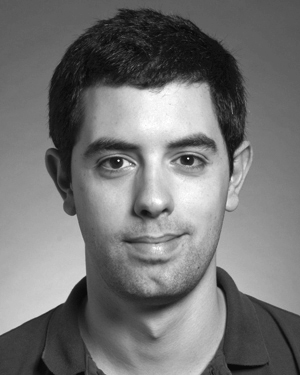}}]{Andre~B.~Reis} is currently working toward the Ph.D. degree in electrical and computer engineering with Carnegie Mellon University (CMU), Pittsburgh, PA, USA, and the University of Aveiro under the CMU-Portugal Program. He received the B.S. and M.Sc. degrees in electronics and telecommunications engineering from the University of Aveiro, Portugal, in 2009, in collaboration with the Eindhoven University of Technology, Netherlands. His current research focuses on infrastructure support systems for vehicular networks in challenging scenarios. He has also published on multimedia Quality of Experience over ad~hoc networks.
\end{IEEEbiography}

\begin{IEEEbiography}
	[{\includegraphics[width=1in,height=1.25in,clip,keepaspectratio]{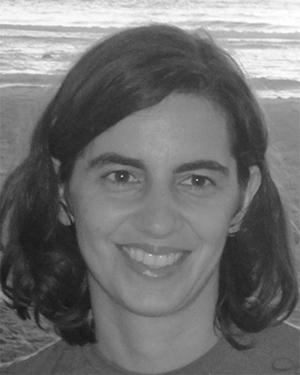}}]{Susana~Sargento} is an Associate Professor with ``Habilitation'' in the University of Aveiro and the Institute of Telecommunications, where she is leading the Network Architectures and Protocols (NAP) group. She has more than 15 years of experience in technical leadership in many national and international projects, and worked closely with telecom operators and OEMs. She has been involved in several FP7 projects, EU Coordinated Support Action 2012-316296 ``FUTURE-CITIES'', national projects, and CMU-Portugal projects. She has been TPC-Chair and organized several international conferences and workshops. She has also been a reviewer of numerous international conferences and journals, such as IEEE Wireless Communications, IEEE Networks, and IEEE Communications. Her main research interests are in the areas of self-organized networks, in ad-hoc and vehicular network mechanisms and protocols, such as routing, mobility, security and delay-tolerant mechanisms, resource management, and content distribution networks. In March 2012, Susana co-founded a vehicular networking company, Veniam, a spin-off of the Universities of Aveiro and Porto, which builds a seamless low-cost vehicle-based internet infrastructure. Susana is the winner of the 2016 EU Prize for Women Innovators.
\end{IEEEbiography}

\vfill\eject
\begin{IEEEbiography}
	[{\includegraphics[width=1in,height=1.25in,clip,keepaspectratio]{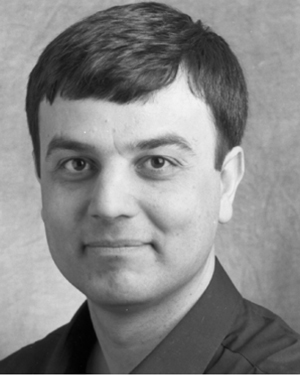}}]{Ozan~K.~Tonguz} is a tenured full professor in the Electrical and Computer Engineering Department of Carnegie Mellon University (CMU). He currently leads substantial research efforts at CMU in the broad areas of telecommunications and networking. He has published about 300 research papers in IEEE journals and conference proceedings in the areas of wireless networking, optical communications, and computer networks. He is the author (with G. Ferrari) of the book \emph{Ad Hoc Wireless Networks: A Communication-Theoretic Perspective (Wiley, 2006)}. He is the inventor of 15 issued or pending patents (12 US patents and 3 international patents). In December 2010, he founded the CMU startup known as Virtual Traffic Lights, LLC, which specializes in providing solutions to acute transportation problems using vehicle-to-vehicle (V2V) and vehicle-to-infrastructure (V2I) communications paradigms. His current research interests include vehicular networks, wireless ad hoc networks, sensor networks, self-organizing networks, artificial intelligence (AI), statistical machine learning, smart grid, bioinformatics, and security. He currently serves or has served as a consultant or expert for several companies, major law firms, and government agencies in the United States, Europe, and Asia.
\end{IEEEbiography}
\vfill

\end{document}